\documentclass[3p,sort&compress]{elsarticle}
\usepackage[T2A]{fontenc}
\usepackage[cp1251]{inputenc}
\usepackage[english,russian]{babel}
\usepackage{graphicx,caption,subcaption}
\usepackage{amsmath}
\usepackage{amssymb}
\usepackage{ulem}
\usepackage{makeidx}
\usepackage{fancyhdr}
\usepackage{lineno}
\usepackage{color}

\begin{document}

\def\figurename{Fig.}
\def\tablename{Table}
\def\refname{References}

{\bf Effect of the electrical double layer on the electrical
conductivity of suspensions}

\bigskip

M. Ya. Sushko\footnote{Corresponding author. E-mail:
mrs@onu.edu.ua} and S. D. Balika

\bigskip

{\it Department of Theoretical Physics and Astronomy, Mechnikov
National University, 2~Dvoryanska St.,
 Odesa 65026, Ukraine}

\bigskip

\bigskip

\section*{Abstract}

We study the role of the electrical double layer (EDL) in the
formation  of the quasistatic electrical conductivity of
suspensions of nanosized particles. A suspension is viewed as a
system of hard-core--penetrable-shell particles. The shells are
electrically inhomogeneous, with a radially symmetrical
conductivity profile. It is assumed that the real microstructure
of the suspension can be reflected in terms of this profile and
also the rule of dominance for overlapping regions that the local
conductivity in the system is determined by the nearest particle.
Using our earlier rigorous results for systems with this
morphology, we derive general integral relations for the desired
conductivity which incorporate the effect of the EDL and make it
possible to look into the contributions from its different parts
and parameters. Specific features, internal consistency, and
flexibility of the model are demonstrated by further elaborating it to describe experimental data for latex suspensions in aqueous
electrolyte solutions with high ionic strength.

\section{Introduction}\label{sec1}
Development of new materials that, on the one hand, possess
desired and controllable electrophysical properties and, on the
other, are relatively cheap to manufacture requires reliable
theoretical methods for evaluation of their relevant
characteristics. Random heterogeneous materials (like composite
solid electrolytes  \cite{Liang1973,Dudney1989,Takahashi1989,Nan1993}, composite polymeric  electrolytes \cite{Knauth2008,Sequeira2010}, or suspensions in electrolyte solutions \cite{Ohshima12}) represent a class of systems that are
already in wide use, but whose effective parameters (the effective
electrical conductivity $\sigma_{\rm eff}$ among those) are very
difficult to analyze theoretically. Two of the challenging issues
in doing so are: (a) modeling of the microstructure of the system,
with consideration for the presence of interface regions and
possible variations of the microstructure with the volume
concentration $c$ of the filler; and (b) electrodynamic
homogenization (finding $\sigma_{\rm eff}$) of the proposed model,
with taking into account many-particle effects.

This work presents a many-particle approach designed to quantitatively describe the effect of the electrical double layer (EDL) on the quasistatic $\sigma_{\rm eff}$  of suspensions. The EDL is a spatial distribution of ions  that forms near a charged particle-base liquid interface. Its structure is  rather complex \cite{Lyklema1998,Delgado2005}, but for our purposes can be considered as consisting of two major parts (see figure~\ref{fig:model}):
\begin{enumerate}
    \item
    The inner part  that comprises the Stern layer (with thickness of
    order of the ionic radius) and the   stagnant
    layer (which is the starting part of the diffuse EDL). The
    electric potential at the boundary between these layers is called
    the diffuse-layer or Stern potential, $\psi$.
    \item The outer part that is the remaining part of the diffuse EDL.
\end{enumerate}

The two parts of the EDL are separated by the slip plane (at a
distance $d^*$ from the interface) the potential at which is the
zeta-potential, $\zeta$. In contrast to $\psi$, $\zeta$ is
considered as a rather well-defined quantity obtainable from
electrophoretic measurements.

The stagnant layer is believed to be hydrodynamically immobile,
but, possibly, electrically conductive. The conductivity of the
outer part is formed by hydrodynamically mobile ions moving under
the electric field in the bulk of the carrying liquid. Because of
thermal  motion of ions, the slip plane between them is in reality
partly spread, not  sharp.

Standard  electrokinetic theories \cite{O'Brien1978,Saville79,OBrien81,DeLacey81}
disregard the presence of the stagnant layer. They  reduce the
problem of finding  $\sigma_{\rm eff}$  to an analysis of a system
of coupled nonlinear differential equations and relations for the
flow field and quantities  related to the ion and electric
potential distributions around a single particle. However, if  the
particle is viewed as a  uniform hard sphere (of radius $a$), the
system of the governing equations (including the Poisson-Boltzmann
equation) is written for the exterior of the sphere, and the
boundary conditions at the particle's surface are those for the
hard sphere, then systematic discrepancies occur between  the
values of $\zeta$ obtained from electrophoretic mobility and
conductivity  calculations
\cite{Zukoski1985}. These discrepancies are explained by a
transport of ions within the Stern layer   \cite{Dukhin1974,Zukoski1986a,Zukoski1986b}. Analysis   \cite{Mangelsdorf1990,Mangelsdorf1998} of
the modifications in the Stern-layer transport theory for two
mechanisms  of Stern-layer adsorption of ions (onto available
surface area and  onto underlying surface charge) and the
resulting boundary conditions shows that, regardless of the
mechanism, the presence of mobile Stern-layer ions causes the
electrophoretic mobility to decrease and the electrical
conductivity to increase, as compared to the case when surface
conduction is absent. The same conclusion is drawn from a cell
model  \cite{Carrique2001} for concentrated
suspensions with hydrodynamic and electric interactions between
particles.

It is important to emphasize that according to its original
definition,  surface conductance is the excess conduction within
the slip plane; in general, it incorporates the effects from the
stagnant layer  \cite{Shilov1970},    in
addition to those from the Stern-layer. Now that the existence of
a hydrodynamically stagnant layer is confirmed by molecular
dynamic simulations  \cite{Lyklema1998},
attempts are necessary to investigate its role in the formation of
electrokinetic parameters and correctly interpret experimental
data. One such an attempt
\cite{Kijlstra1992,Kijlstra1993}, free of in-depth
detailing the structure of the stagnant layer, represents an
extension  of  the thin-double-layer theory
\cite{Fixman1980,Fixman1983} for dilute sols of
spherical particles.

Incorporating the stagnant layer (or an analogue of it) into the
model for  $\sigma_{\rm eff}$ seems very promising. In what
follows, we suggest that the factors behind the behavior of
$\sigma_{\rm eff}$ with $c$ can be illustrated by this schematic
picture:
\begin{enumerate}
    \item The effective conductivity $\sigma_{\rm eff}$ at low $c$ is
    first of all a result of formation of conducting paths in the bulk
    of the suspension. These paths are formed by the mobile ions
    beyond the slip plane, that is, the outer part  of the diffuse
    EDL.
    \item The potential $\zeta$ at and the conductivity
    distribution within the slip plane result from the surface
    conduction processes occurring within the inner part of the EDL at
    a given $\kappa a$ ($\kappa^{-1}$ is the Debye length).
    \item The
    location of the slip plane is controlled by the thickness $d^*$ of
    the inner part of the EDL. In view of an ionic-size thickness of
    the Stern layer,  this is actually the thickness of the stagnant
    layer (or an analogue of it).
    \item The properties of the stagnant
    layer are system-dependent. As a result, even for suspensions with
    equal $\zeta$ and $\kappa a$, $\sigma_{\rm eff}$ is not a
    universal function of  $c$, but also depends on the relative thickness
    $u^*=d^*/a$ of the inner part of the EDL. Data obtained by several
    electrokinetic techniques must be available to reliably estimate
    the governing parameters.
    \item For given $\zeta$, $\kappa a$, and
    $ u^*$, $\sigma_{\rm eff}$ changes in response to transformations
    of the system of conducting paths in the bulk, which include (at
    sufficiently high $c$) overlappings of the outer parts of the
    EDLs. The values of $c$ at which the inner parts of the EDLs start
    overlapping are hardly achievable for real suspensions.
\end{enumerate}

Viewing a suspension as a system of hard-core--penetrable-shell
particles  \cite{Sushko2016}, generalizing this
model to the case of inhomogeneous shells
\cite{Sushko2019jml,Sushko2019pre},  and using the
compact-group approach  \cite{Sushko2007,Sushko2017} for electrodynamic homogenization, we can
integrate these factors and their effects into a single model to
derive and analyze  rigorous integral relations for $\sigma_{\rm
    eff}$ of suspensions with different conductive properties of the
stagnant layer.

The subsequent presentation of our results is organized as
follows. The basics of the core-shell model and compact-group
approach are outlined in section \ref{sec2}. The elaboration of
the model for suspensions and the derivation of the indicated
relations for $\sigma_{\rm eff}$ are given in section \ref{sec3}.
Applications of the theory to real electrolyte-based suspensions
are discussed in section \ref{sec4}. The major results of the work
are summarized in section \ref{sec5}.

\section{Model and method for finding $\sigma_{\rm eff}$}\label{sec2}
Describing the model under consideration and the method for its
homogenization, we omit numerous technical details and
justifications; the interested reader will find them in the above
mentioned articles
\cite{Sushko2016,Sushko2019jml,Sushko2019pre,Sushko2007,Sushko2017} by the authors.

\subsection{Microstructure of the model system}\label{subsec21}
\begin{figure}[bth]
    \centering
    \includegraphics[width=60mm]{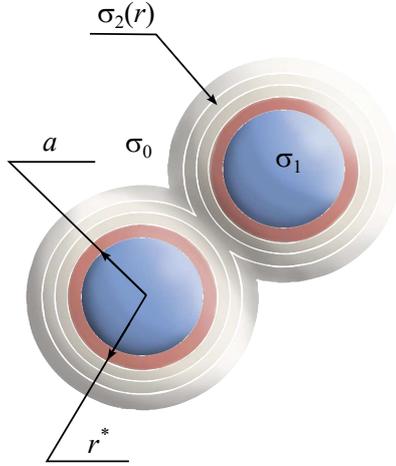}
    \caption{\label{fig:model}
        Hard-core--penetrable-shell particles
        under consideration. Each shell consists of two concentric parts,
        separated by a spherical surface of radius $r^*$ (at relative
        distance $ u^* =(r^*-a)/a$ from the core). The inner and outer
        shells are associated with the inner and outer parts of the EDL,
        and the separating surface is associated with the slip plane at
        $\zeta$-potential. The radius of the outer shell, shown as finite
        for illustrative purposes, is taken to tend to infinity in the
        final formula.
        The complex permittivity of each constituent
        [including the fictitious matrix($\hat{\varepsilon}_{\rm f}$)   and the homogenized system ($\hat {\varepsilon}_{\rm eff}$), both
        not shown]  is taken in the form $\hat{\varepsilon}=\varepsilon
        +{\rm i} \sigma/e_0 \omega$, where $\varepsilon $ and $\sigma$ are respectively
        the quasistatic dielectric constant and electrical conductivity of
        the constituent, $\omega$ is the angular frequency
        of probing radiation, $e_0$ is the electric constant, and ${
            \rm i}$ is the imaginary unit.
        The resulting relation for $\sigma_{\rm
            eff} $ involves the conductivities $\sigma_1$, $\sigma_2(r)$, and
        $\sigma_0$ of the core, shell, and base liquid, respectively. The
        variable  $u=(r-a)/a$ defines the relative distance from the
        surface of the core to the point of interest.
    }
\end{figure}

We assume that the electrical properties of a real suspension are
equivalent to those of a model system that is made up by embedding
hard-core--penetrable-shell particles  into a fictitious uniform
matrix with a complex permittivity $\hat{\varepsilon}_{\rm f}$
(see figure~\ref{fig:model}). The particles are considered to be
stationary. This is a typical approximation  in electrical
conductivity problems, if the particles are sufficiently massive
and  alternating probing fields (of even low frequencies) are
used. Experimentally its validity is controlled by varying the
measurement frequency to make sure that the  measurement results
are independent of it. The microstructure of this model system is
described in terms of the low-frequency complex permittivity
distributions for its constituents. Each core is a uniform hard
sphere of radius $a$ and complex permittivity
$\hat{\varepsilon}_1$; it is  associated with a real particle.
Each shell is isotropic, but electrically inhomogeneous in the
radial direction. Its complex permittivity profile is a piecewise
continuous function $\hat{\varepsilon}_2= \hat{\varepsilon}_2 (r)$
that approaches  the complex permittivity $\hat{\varepsilon}_0$ of
the base liquid as $r \to \infty$, $r$ being the distance from the
center of the sphere to the point of interest. The explicit
expression for $\hat{\varepsilon}_2 (r)$ is modeled so as to
account for different mechanisms contributing to the electrical
characteristics of the real suspension at different values of $c$.
For this purpose, the theory is complemented by a rule of
dominance  imposed on the electrical properties of overlapping
regions which states: (a) the electrical properties of hard cores
dominate over those of the shells and (b) those of closer (to the
surface of a given core) parts of the shells over those of farther
parts. This rule is necessary to define  the electrical
microstructure of the  suspension uniquely, with consideration for
its variations with $c$ (see \cite{Sushko2019pre,Sushko2019jml}).
Physically, this rule means that the local value of the complex
permittivity in the system is determined by the distance from the
point of interest to the closest particle. The outermost part  of
${\hat\varepsilon_2 (r)}$ accounts for  the electrical properties
of the suspending liquid  and  its contribution to  $\sigma_{\rm
eff }$.

\subsection{Basics of the compact group approach} \label{subsec22}
The quasistatic response of the model system to probing radiation
is analyzed using the notion of compact groups of inhomogeneities.
These groups are defined as macroscopic regions that are much
smaller than the wavelength $\lambda$ of the probing field in the
medium, yet large enough to retain the properties of the entire
system. The model system is viewed as a collection of such
regions, with the complex permittivity distribution in it given by
\begin{eqnarray}\label{profile}
    \hat{\epsilon}(\bf{r}) = \hat{\varepsilon}_{\rm f} + \delta
    \hat{\varepsilon}(\bf{r}),
\end{eqnarray}
where $\delta \hat{\varepsilon}(\bf{r})$ is the contribution from
the compact group located at point $\bf{r}$ and comprising both
particles and the base liquid. Since compact groups are point-like with respect
to $\lambda$, their contributions to the average complex electric
field $\langle{\rm {\bf {E}}}\rangle $ and average complex electric current
$\langle{\rm {\bf {J}}}\rangle$ are formed by those values of the
coordinates where the inner propagators in the corresponding
iterative series reveal singular behavior. Moreover, for a
macroscopically isotropic and homogeneous system it is only the
$\delta$-function singularities in these propagators that give
non-zero contributions to $\langle{\rm {\bf {E}}}\rangle $ and
$\langle{\rm {\bf {J}}}\rangle$. This fact allows us to single out
the compact group contributions from the iterative series to
express $\langle{\bf {E}}\rangle $ and  $\langle{\rm {\bf
        {J}}}\rangle$  in terms of the moments of $\delta
\hat{\varepsilon}(\bf{r})$ without employing any specific
assumptions about multiple reemissions and many-particle
correlations inside compact groups:
\begin{eqnarray} \label{eq:field} \langle{\rm {\bf  {E}}}\rangle =
    {\left[ {1 + {\sum\limits_{s = 1}^{\infty}  {\left( { -
                        {\frac{{1}}{{3\hat{\varepsilon}_{\rm f} }} }} \right)^{s} \langle
                    {{\mathop {\left( {\delta \hat{\varepsilon} ({\rm {\bf r}})}
                                \right)^{s}}}}} }}\rangle \right]}\,{\rm {\bf E}}_{0},
\end{eqnarray}
\begin{eqnarray}\label{eq:current}
    \langle{\rm {\bf {J}}}\rangle = -{\rm i} \omega e_0
    \hat{\varepsilon}_{\rm f}{\left[ { 1 - 2 {\sum\limits_{s =
                    1}^{\infty}  {\left( { - {\frac{{1}}{{3\hat{\varepsilon}_{\rm f}
                    }}}} \right)^{s}{\langle {\mathop {\left( {\delta
                                    \hat{\varepsilon} ({\rm {\bf r}})} \right)^{s}}}}\rangle} }}
        \right]}\,{\rm {\bf E}}_{0},
\end{eqnarray}
where ${\rm {\bf E}}_0$ is the amplitude of the probing field in the fictitious
matrix.  The effective permittivity $\hat{\varepsilon}_{\rm{eff}}$
is defined as the proportionality coefficient in the relation
\begin{eqnarray}
    \label{eq:effcomplex} \langle {\bf{J}} ({\bf{r}})\rangle = -{\rm i} \omega e_0 \langle \hat{\varepsilon} ({\bf{r}}) {\bf{E}}
    ({\bf{r}}) \rangle = -{\rm i}\omega e_0  \hat{\varepsilon}_{\rm
        eff} \langle {\bf{E}} ({\bf{r}}) \rangle .
\end{eqnarray}

In his pioneering work \cite{Maxwell1873}, Maxwell calculated the
effective electrical conductivity for a system of uniform hard
spheres by comparing  the large-distance asymptotic expressions
for the electric potentials produced by a single such a sphere and
a large sphere composed of many such spheres, assuming no
electromagnetic interaction among them. The compact group approach
is actually a rigorous and  technically-different implementation
of this idea by using the multiple scattering and generalized
function theories for a system of  particles whose electromagnetic
interactions lead  to the formation of the mean electric field
(\ref{eq:field}) in the system. The moments $ \langle{\left(
{\delta
        \hat{\varepsilon} ({\rm {\bf r}})} \right)^{s}} \rangle$,
$s\geq 1$, of
${\delta
    \hat{\varepsilon} ({\rm {\bf r}})}$
account for the effects of reemissions of order  $s$ within
compact groups ('large spheres'). Maxwell's formula is obtained
from (\ref{eq:field})  and (\ref{eq:current}) under the conditions
that the  particles are hard, $\hat{\varepsilon}_{\rm f} =
\hat{\varepsilon}_0$, and $\omega\to 0$.

The concept of the homogenized medium for a given system is
physically meaningful if (a)  the values of
$\hat{\varepsilon}_{\rm{eff}}$ obtainable by using different
operational definitions for $\hat{\varepsilon}_{\rm{eff}}$ are the
same, and (b) the standard boundary conditions of electrodynamics
remain valid for all constituents of the system and the
homogenized medium as well. The compact group approach is
compatible with these requirements if
\begin{eqnarray}
    \label{eq:closure} \hat{\varepsilon}_{\rm f}=
    \hat{\varepsilon}_{\rm eff}.
\end{eqnarray}
In other words, it is an internally-closed homogenization
procedure of the Bruggeman type.

In view of (\ref{eq:field})--(\ref{eq:closure}), the equation
for $\hat{\varepsilon}_{\rm{eff}}$ takes the form
\begin{eqnarray} \label{eq:generalpermit}
    \sum\limits_{s = 1}^{\infty}  \left( { -
        {\frac{{1}}{{3\hat{\varepsilon}_{\rm f} }}}} \right)^{s}{\langle
        {\mathop {\left( {\delta \hat{\varepsilon} ({\rm {\bf r}})}
                \right)^{s}}}}\rangle =0.
\end{eqnarray}
The equation for $\sigma_{\rm eff} $ is obtained by passing to the
static limit $\omega \to 0$ in this relation.

\subsection{General relation for $\sigma_{\rm eff}$ of the model
    system} \label{subsec23}

Equation (\ref{eq:generalpermit}) holds regardless of the specific
form of the expression for  $\delta \hat{\varepsilon}(\bf{r})$. To
advance,  this expression must be constructed now for the model
system.
Using the indicator   (or characteristic) function, defined for a region   $\Omega$ as
\[\Pi _\Omega \left(\rm \bf r \right) = \left\{ \begin{array}{l}
    1\,{\rm{if}}\,{ {\rm \bf r}} \in \Omega \\
    0\,{\rm{if}}\,{ {\rm \bf r}} \notin \Omega
\end{array} \right.\]
this can be done as follows.

Partitioning each shell into very thin concentric  spherical layers with outer
radii $r_j$ ($> a\equiv r_0$) and complex permittivities
$\hat{\varepsilon}_{2,j} $, we can write down $\delta
\hat{\varepsilon}(\bf{r})$ as
\begin{eqnarray}\label{compgroupper}
    \delta\hat{\varepsilon} ({\bf r})= \left(
    \hat{\varepsilon}_1-\hat{\varepsilon}_{\rm eff} \right)\,
    \Pi_1({\bf r}) + \sum\limits_{j\geq 1}
    \left(\hat{\varepsilon}_{2,j} -\hat{\varepsilon}_{\rm eff}
    \right)\, \tilde {\Pi}_{2,j}({\bf r}),
\end{eqnarray}
where $\tilde {\Pi}_{2,j}({\bf r}) = \Pi_{2,j}({\bf r}) -
\Pi_{2,j-1}({\bf r})$, with $\Pi_{2,j}({\bf r})$ ($ \Pi_{2,0}({\bf
    r}) \equiv \Pi_1({\bf r})$) being the characteristic functions of
the subsystems of balls of radii $r_j$. Then, due to the
properties of characteristic functions and the orthogonality of
any two functions from the set $\{ \Pi_1, \tilde {\Pi}_{2,j} \} $,
we find
\begin{eqnarray}\label{moments}
    \langle\left(\delta\hat{\varepsilon}({\bf r})\right)^s\rangle =
    c\left(\hat{\varepsilon}_1-\hat{\varepsilon}_{\rm eff}\right)^s
    +\sum\limits_{j\geq
        1}\left[\phi(c,u_j)-\phi(c,u_{j-1})\right]\left(\hat{\varepsilon}_{2.j}
    -\hat{\varepsilon}_{\rm eff}\right)^s,
\end{eqnarray}
where $\phi (c, u_j) = \langle \Pi_{2,j}({\bf r}) \rangle$ is
the volume concentration of hard-core--penetrable-shell particles
whose shells have relative thickness $u_j$. Estimates for $\phi
(c, u)$ are obtained by statistical physics calculations  (see, for instance, \cite{Rikvold85,Rikvold85col}).

Passing  in the moments (\ref{moments}) to integration, substituting
them into (\ref{eq:generalpermit}), carrying out the
summation, and, finally, passing to the static limit, we obtain
the general equation for  $\sigma_{\rm eff}$:
\begin{eqnarray} \label{genEqn}
    c\,\frac{\sigma_1 -\sigma_{\rm
            eff}}{2\sigma_{\rm eff}+\sigma_1}
    +\int\limits_0^{\infty}\frac{\partial \phi(c,u)}{\partial
        u}\frac{\sigma_2 (u) - \sigma_{\rm eff}}{2\sigma_{\rm eff}+\sigma_2
        (u)}{\rm d}u =0 .
\end{eqnarray}
This equation is rigorous in the static limit and serves as the
starting point for further analysis of suspensions. In the case of
finite-thickness shells, it reproduces our earlier equation
\cite{Sushko2019pre} which has proven to
be efficient in describing composite solid and composite polymeric
electrolytes.

\section {Elaboration of the model for suspensions}\label{sec3}

Equation (\ref{genEqn}) is quite general. Now, we intend to add
more details to the model to account for specific features of the
EDL and the effects of its different parts on $\sigma_{\rm eff}$
of suspensions. In doing so, we focus on the situations where the
number of fitting parameters in the model can be reduced to the
least possible.

\subsection {Accounting for the Stern layer}\label{subsec31}
As was indicated, ion transport in the Stern layer can change the
conductivity of the particle. Taking into account that the
thickness of this layer is comparable with the ion size and,
macroscopically, it is located at $u=0$, we suggest that this
effect can be incorporated into the theory by redefining
$\sigma_1$ to be the conductivity  that the particle has in the suspension, not before being embedded into it. Similar alteration of the conductive properties of
constituents in the course of combining them into a system is
observed, for instance, in composite polymeric electrolytes
\cite{Sushko2019jml}.

The other effects of the Stern layer---on the adjacent stagnant
layer and the potential distribution near the particle---are
actually incorporated via the parameters (conductivity and
thickness) of the stagnant layer and $\zeta$-potential, as
discussed below.

\subsection {Accounting for the stagnant layer}\label{subsec32}

Due to a specific structure of the integrand, three particular
situations can be analyzed in-depth based only on very general
assumptions about the behavior of $\sigma_2(u)$ in the stagnant
layer, by which the inner part  $u\in (0,u^*)$ of the diffuse EDL
is meant:
\begin{enumerate}
    \item {\it Highly-conductive stagnant layer}: $\sigma_2 (u) \gg
    \sigma_{\rm eff}$ for $u \in (0, u^*)$.

    In this case, splitting the integral in (\ref{genEqn}) into one
    over the region  $u \in (0, u^*)$ and the other over the region $u
    \in (u^*,\infty)$ gives
    \begin{center}
        \begin{eqnarray} \label{case1a}
            c\,\frac{\sigma_1 -\sigma_{\rm
                    eff}}{2\sigma_{\rm eff}+\sigma_1}
            + F(c, u^*)
            +\int\limits_{u^*}^{\infty}\frac{\partial \phi(c,u)}{\partial
                u}\frac{\sigma_2 (u) - \sigma_{\rm eff}}{2\sigma_{\rm eff}+\sigma_2
                (u)}{\rm d}u =0,
        \end{eqnarray}
    \end{center}
    where
    \begin{eqnarray} \label{Fcase1}
        F(c, u^*) =\int\limits_0^{u^*}\frac{\partial \phi(c,u)}{\partial u}{\rm d}u = \phi(c, u^*) - c,  \nonumber
    \end{eqnarray}
    for $\phi(c, 0)=c$. Then, resolving the fraction in the integrand
    into partial fractions, adding and subtracting the term
    \begin{eqnarray}
        3\sigma_{\rm eff} \int\limits_{u^*}^{\infty}\frac{\partial
            \phi(c,u)}{\partial u} \frac{1}{2\sigma_{\rm eff}+\sigma_0 } {\rm d}u,
        \nonumber
    \end{eqnarray}
    and taking into account that
    \begin{eqnarray}
        \int\limits_{u^*}^\infty \frac{\partial \phi(c,u)}{\partial u}{\rm d}u = 1- \phi(c, u^*), \nonumber
    \end{eqnarray}
    we obtain
    \begin{eqnarray} \label{mainEq}
        \left[1-\phi\left(c,u^*\right)\right]\frac{\sigma_0 - \sigma_{\rm eff}}{2\sigma_{\rm eff}+\sigma_0}  + c\,\frac{\sigma_1 -\sigma_{\rm eff}}{2\sigma_{\rm eff}+\sigma_1} +  F(c, u^*) \nonumber \\ - 3\sigma_{\rm eff} \int\limits_{u^*}^{\infty}\frac{\partial
            \phi(c,u)}{\partial u}\left[\frac{1}{2\sigma_{\rm eff}+\sigma_2 (u)}-\frac{1}{2\sigma_{\rm eff}+\sigma_0 }\right] {\rm d}u=0.
    \end{eqnarray}

    In particular, for nonconducting particles ($\sigma_1\to 0$) (\ref{mainEq}) reduces to
    \begin{eqnarray}   \label{main0Eq}
        \left[ 1-\phi(c,u^*)\right]\frac{\sigma_0 - \sigma_{\rm eff}}{2\sigma_{\rm eff}+\sigma_0} +  F_0(c, u^*)\nonumber\\
        -3\sigma_{\rm eff} \int\limits_{u^*}^{\infty}\frac{\partial
            \phi(c,u)}{\partial u}\left[\frac{1}{2\sigma_{\rm eff}+\sigma_2
            (u)}-\frac{1}{2\sigma_{\rm eff}+\sigma_0 }\right] {\rm d}u=0
    \end{eqnarray}
    with
    \begin{eqnarray}   \label{F0case1}
        F_0(c, u^*) = \phi(c, u^*)  - \frac{3}{2}c .
    \end{eqnarray}

    \item {\it Low-conductive stagnant layer}: $\sigma_2 (u) \ll \sigma_{\rm eff}$ for $u \in (0, u^*)$.

    Proceeding in the same way, we arrive at (\ref{mainEq})
    and (\ref{main0Eq}) again, but with
    \begin{eqnarray} \label{FF0case2}
        F(c, u^*) =- \frac{1}{2} \left[\phi(c, u^*) - c\right], \quad
        F_0(c, u^*) = - \frac{1}{2}\phi(c, u^*).
    \end{eqnarray}

    \item {\it Conductivity of the stagnant layer is comparable with the effective one}:
    $\sigma_2 (u)\approx \sigma_{\rm eff}$ for $u \in (0, u^*)$.

    Now, assuming the contribution from the integral over $u \in (0, u^*)$ to be
    negligible as compared to that over $u > u^*$, we have
    (\ref{mainEq}) and (\ref{main0Eq}) with
    \begin{eqnarray} \label{FF0case3}
        F=0, \quad F_0(c) = - \frac{1}{2} c.
    \end{eqnarray}
    \end {enumerate}

    Note that the physical meaning of each addend in
    (\ref{mainEq}) is clear. The first one is the contribution from
    the base liquid. The second one describes the contribution from
    the particles whose conductivities incorporate the effect of ionic
    transport in the Stern layer. The third one is the contribution
    from the inner part of the diffuse EDL. And the fourth one
    describes the effect of the outer part of the diffuse EDL. Of
    course, all these contributions are interrelated.

    \subsection {Accounting for the outer part of the EDL}\label{subsec33}

    Beyond the slip plane, the conductivity is formed by mobile ions
    in the diffuse EDL. Let  $z_a$, $\mu_a$, and $n_a$  be
    respectively the charge numbers, mobilities, and average number
    densities of the ions in the carrying liquid. Assuming the
    Boltzmann distribution to be valid for the region $u>u^*$ about a
    given particle, we model the one-particle profile $\sigma_2(u)$
    for $u>u^*$ as
    \begin{eqnarray} \label{sigma2beyond}
        \sigma_2(u) = e \sum_a |z_a|\, \mu_a n_a  e^{-z_a y(u)},
    \end{eqnarray}
    where $y(u)= e\varphi(u)/k_{\rm B}T$, $\varphi(u)$ is the electric
    potential distribution for $u>u^*$, such that
    $\varphi(u^*)=\zeta$, $ k_{\rm B}$ is the Boltzmann constant, $T$
    is the temperature, and $e$ is the elementary charge.
    Neither  $u^*$ nor $y(u)$ is
    supposed to be extrapolated to the surface of the particle. In general, $y(u)$ is unknown---since it depends on the locations
    of the other particles as well, finding it for arbitrary $c$ is a
    challenging many-particle problem. Yet in situations where the
    overlapping of the EDLs can be ignored, we expect well-known
    one-particle solutions to be applicable  for $y(u)$.

    \section{Application to electrolyte-based suspensions}\label{sec4}

    To put our model to the test and elucidate the role  that  the diffuse EDL can play  in the formation
    of $\sigma_{\rm eff}$, we consider
    here  a suspension of nonconducting particles  in an aqueous
    1:1 electrolyte solution  with high ionic strength.
    Extensive experimental data for $\sigma_{\rm eff }$ of such suspensions are given  in  \cite{Zukoski1985}.
    In order to reduce the number of fitting parameters to the least number possible, we  also assume that (a)  the
    effect of the Stern layer on the conductivity of the particles is
    negligible, so  $\sigma_1=0$; and (b) the ion mobilities in the EDL beyond the slip plane are equal to those in the bulk of the suspending liquid.
    It turns out that under these assumptions, experimental data
    \cite{Zukoski1985}
    can be recovered using only a single fitting parameter,  the relative thickness $u^*$ of the stagnant layer.

    The models  with  an adjustable $\sigma_1$   and   the ion mobilities whose  values near the  surfaces of  particles (say, due to the convective contribution \cite{Bikerman1940}) and those in the bulk  are  in general  different  are,  of course,  more flexible and   include  the case under consideration. They are also encompassed by our general formalism. The study of them is a matter for further analysis which, however, goes beyond the scope of this work.

    \subsection{Working equations}\label{sec41}
    The fact that $\kappa a \gg 1$  means that, first, overlapping of
    the EDLs is negligible up to rather high $c$; and, second, the
    curvature of the EDL is very small, which implies that the
    Gouy-Chapman solution for planar surfaces can be used to describe
    the radial distribution of the electric potential in the EDL
    beyond the slip plane:
    \begin{eqnarray}\label{eqGouy}
        y(u) = 2\ln \left[ {\frac{{1 +
                    \gamma {e^{ - \kappa a \left(u-u^*\right)}}}}{{1 - \gamma {e^{
                            -\kappa a \left(u-u^*\right)}}}}} \right], \quad
        \gamma  = \tanh
        \left( {\frac{{{y_\zeta}}}{4}} \right), \quad
        y_\zeta=\frac{{{e\zeta}}}{k_{\rm B}T} .
    \end{eqnarray}
    The corresponding distribution of the electrical conductivity is
    \begin{eqnarray}\label{eqGouyConuctivity}
        \sigma_2(u) = e\mu_+n_0 e^{-y(u)} + e\mu_-n_0e^{y(u)}, \quad
        u>u^*,
    \end{eqnarray}
    where $\mu_+$ and $\mu_-$ are respectively the cation and anion mobilities (assumed
    to be independent of the electric field), and $n_0 $ is the
    average number density of ions of each type. As $u\to\infty$,
    $\sigma_2(u)$ tends to $\sigma_0 = e\mu_+n_0 + e\mu_-n_0$, the
    conductivity of the carrying liquid.

\begin{figure}[bth]
        \centering
        \includegraphics[width=75mm]{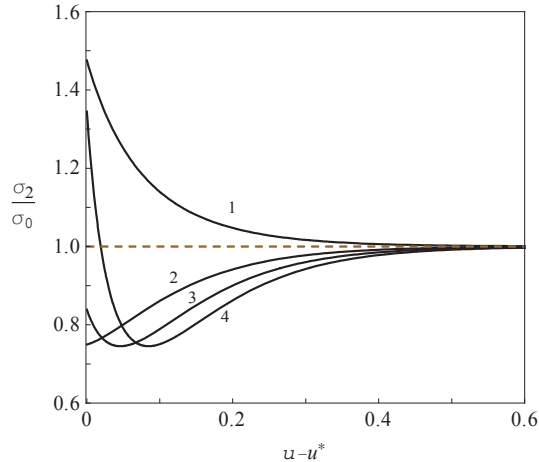}
        \caption{\label{fig:condtypes}
            Normalized conductivity
            $x_2(u) = \sigma_2(u)/\sigma _0$
            as a function of
            $u-u^*$
            for $ u\geq u^*$,
            according  to (\ref{eqGouy}) and (\ref{eqGouyConuctivity}) at
            $\kappa a =10$, $m =5$, and $y_\zeta = - 0.5$, 0.7, 1,3, and 2
            (curves 1 to 4, respectively).
            The plots for $0<m<1$ and the
            opposite values of $ y_\zeta $ look similar. The local minimum
            $ x_2(u_0)=2 m^{1/2}/(m+1) <1$
            at
            $ u_0=u^*+(\kappa a)^{-1} \ln |\gamma\, (m^{1/4}+1)/ (m^{1/4}-1)|$
            appears beyond the
            slip plane if
            $|\gamma|>|m^{1/4}-1|/(m^{1/4}+1)$.}
    \end{figure}

    Denote $m\equiv\mu_+/\mu_-$. Simple analysis reveals that provided
    (a) $m>1$ for $\zeta>0$ or (b) $0<m<1$ for $\zeta<0$, there are
    intervals of $u$ where $\sigma_2(u) <  \sigma_0$ (see
    figure~\ref{fig:condtypes}). The volume fraction of such regions
    increases as $c$ is increased, which can cause  $\sigma_{\rm eff
    }$ to decrease even in the case of highly conductive stagnant
    layers. On the other hand, the condition $\sigma_2(u) > \sigma_0$
    does not assure that $ \sigma_{\rm eff}$ is an increasing function
    of $c$.

    In what follows, we change to the normalized conductivities
    $x_2(u) = \sigma_2(u)/\sigma_0$ and $x_{\rm eff}  = \sigma_{\rm eff}/\sigma_0$.  According to (\ref{main0Eq}) and
    (\ref{F0case1}), the  suspension conductivity increment
    $\sigma_0^{-1} {\partial \sigma_{\rm eff}}/{\partial c}$ ($c\to
    0$), that is, the normalized rate of change of $\sigma_{\rm eff }$
    with respect to $c$ for a diluted suspension, is given (in the
    case of highly conductive stagnant layers) by
    \begin{eqnarray}\label{eq:rate}
        \frac{\partial x_{\rm eff}}{\partial c}=3 (1+u^*)^3 -\frac{9}{2}
        +9 \int_{u^*}^{\infty} (1+u)^2 \frac{x_2(u) - 1}{2+x_2(u)}  {\rm d}u .
    \end{eqnarray}
    Also, introduce the second radial moment of the deviation of the
    normalized conductivity distribution beyond the slip plane from
    the conductivity of the carrying liquid:
    \begin{eqnarray}\label{eq:2RadMoment}
        m_2(r^*) = \int_{r^*}^{\infty} r^2\left[x_2(r)-1\right] {\rm d}r =
        a^3
        \int_{u^*}^{\infty} (1+u)^2\left[x_2(u)-1\right]{\rm d}u .
    \end{eqnarray}

    With the use of these two quantities, the behavior of $\sigma_{\rm
        eff}$ for the situations in figure~\ref{fig:condtypes} can be
    quantified as follows:

    \begin{enumerate}
        \item $x_2(u)<1$ {\it  for $u >  u^*$  (curves 2 and 3 in figure~\ref{fig:condtypes})}
        \subitem The conductivity $ \sigma_{\rm eff} $ decreases, $\partial x_{\rm eff}/{\partial c}<0$, provided
        \begin{eqnarray}
            \label{eq:condDecrease}
            \int_{u^*}^{\infty} (1+u)^2\frac{ x_2 (u)-1}{2+x_2 (u)} {\rm d}u <
            \frac{1}{2}- \frac{1}{3} (1+u^*)^3.
        \end{eqnarray}

        The integral is negative, so this inequality holds for $u^*$ up to
        a value of $ (3/2)^{1/3}-1\simeq 0.145 $, at least. For greater
        values of $u^*$, a further system-dependent analysis is needed.
        \subitem
        Note that since now
        \begin{eqnarray*}
            \int_{u^*}^{\infty} (1+u)^2 \frac{ x_2 (u)-1}{2+x_2 (u)} {\rm d}u \geq
            \frac {1}{2}\, \int_{u^*}^{\infty} (1+u)^2\left[x_2(u)-1\right]{\rm d}u,
        \end{eqnarray*}
        a necessary condition for $\sigma_{\rm eff}$ to decrease can be
        written as
        \begin{eqnarray}\label{eq:condDecrease2RadMoment}
            \frac{m_2(r^*)}{a^3} < 1 - \frac{2}{3} \frac{ r^{*3}}{ a^3}.
        \end{eqnarray}
        \subitem
        Ignoring  the existence of  the diffuse EDL leads to the
        Maxwell-Garnett result
        \begin{eqnarray*}
            \frac{\partial x_{\rm eff}}{\partial c}
            \xrightarrow[c\to 0]{} -\frac{3}{2}.
        \end{eqnarray*}

        \item $x_2(u) >1$ {\it for $u >  u^*$  (curve 1 in
            figure~\ref{fig:condtypes})}
        \subitem
        The conductivity $ \sigma_{\rm eff} $ increases, $\partial x_{\rm
            eff}/{\partial c}>0$, if
        \begin{eqnarray}\label{eq:condIncrease}
            \int_{u^*}^{\infty} (1+u)^2\frac{ x_2 (u)-1}{2+x_2 (u)} {\rm d}u >
            \frac{1}{2}- \frac{1}{3} (1+u^*)^3.
        \end{eqnarray}
        The integral is now positive, so the inequality definitely holds
        for $u^* > (3/2)^{1/3}-1\simeq 0.145 $. This estimate agrees with
        our earlier result for a model suspension of hard-core particles
        surrounded by highly-conductive uniform penetrable shells  \cite{Sushko2016}. For smaller values of $ u^*$, an
        additional analysis is required.

        \subitem In the case under consideration,
        $$
        \int_{u^*}^{\infty} (1+u)^2 \frac{ x_2 (u)-1}{2+x_2 (u)} {\rm d}u \leq
        \frac {1}{3}\, \int_{u^*}^{\infty} (1+u)^2\left[x_2(u)-1\right]
        {\rm d}u .
        $$
        If $\sigma_{\rm eff}$ increases, then, in view of the two previous
        equations,
        \begin{eqnarray}\label{eq:condIncrease2RadMoment}
            \frac{m_2(r^*)}{a^3} > \frac{3}{2}- \frac{ r^{*3}}{ a^3}.
        \end{eqnarray}
        Physically, (\ref{eq:condIncrease2RadMoment}) is a necessary
        condition for $ \sigma_{\rm eff}$ to increase with $c$ that is
        imposed on $ m_2(r^*)$. The value of this moment can be increased
        by (a) increasing $|\zeta|$, (b) decreasing $\kappa a$, or (c)
        combining both ways.

        \item {\it $x_2(u) >1$ for $u \in (u^*, u^{**}), but \, x_2(u) <1$ for $u >  u^{**}$ (curve 4 in figure~\ref{fig:condtypes})}
        \subitem The conductivity $ \sigma_{\rm eff} $ decreases if
        inequality (\ref {eq:condDecrease}) holds. Split the integral into
        one over $(u^*,u^{**})$ and the other over $(u^{**}, \infty)$, where
        $u^{**}$ is the solution to the equation $ x_2(u^{**})=1$. Since
        the latter integral is negative, omitting it gives a sufficient
        condition for $\sigma_{\rm eff}$ to decrease:
        \begin{eqnarray}\label{eq:condDecrease2}
            \int_{u^*}^{u^{**}} (1+u)^2\frac{ x_2 (u)-1}{2+x_2 (u)} {\rm d}u <
            \frac{1}{2}- \frac{1}{3} (1+u^*)^3 .
        \end{eqnarray}
        For  (\ref{eqGouy}) and (\ref{eqGouyConuctivity}), $u^{**} =
        u^{*} + (\kappa a)^{-1} \ln |\gamma\, (m^{1/2}+1)/
        (m^{1/2}-1)|$.
    \end{enumerate}

    To complete the model, we use the following estimate for $\phi
    (c, u)$  \cite{Rikvold85col} in our
    calculations:
      \begin{eqnarray} \label{eq:volumeconcpenetrable}
        \varphi (c, u) = 1- (1 - c) \exp\left[
        - \frac{((1+u)^3 -1)c}{1-c} \right]
        \exp\left\{-\frac{3(1+u)^3c^{2}}{2(1-c)^3}
        \right.\nonumber \\
        \times \left.  \left[2 - \frac{3}{1+u} + \frac{1}{(1+u)^3} -
        \left(\frac{3}{1+u} - \frac{6}{(1+u)^2} + \frac{3}{(a+u)^3}
        \right) c  \right] \right\}.
    \end{eqnarray}
    For diluted suspensions $\left( c \to 0 \right)$,
    \begin{eqnarray}
        \phi (c, u) = 1 - \exp\left[-\left( 1 + u \right)^3  c \right]
        + O \left( c^2 \right).
    \end{eqnarray}

    Combined together,  (\ref{main0Eq})--(\ref{FF0case3}),
    (\ref{eqGouy}), (\ref{eqGouyConuctivity}), and
    (\ref{eq:volumeconcpenetrable}) make up closed models that differ
    by the values of the conductivity of the stagnant layer. We are
    now in a position to contrast their results with experiment.

    \subsection{Comparison with experiment}\label{sec42}

\begin{figure}[bth]
    \centering
    \includegraphics[width=75mm]{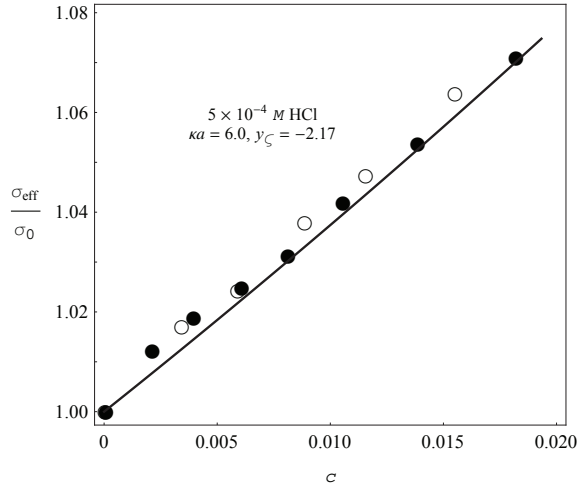}
    \caption{\label{fig:confuctivity2} $\sigma_{\rm eff}$ as a
        function of $c$ for suspensions of Latex A particles
        in $5\times 10 ^{-4}$ $M$ HCl \cite{Zukoski1985}. Solid line: our model results for
        reported $\kappa a$ and $\zeta$ and the
        fitting value of $u^* = 0.28$ ($d^*=1.68 \, \kappa^{-1}$). The
        corresponding theoretical estimate $\partial x_{\rm eff}/\partial
        c = 3.61 $ is in agreement with experimental data $3.57 \pm 0.06$.
        Condition (\ref{eq:condIncrease2RadMoment}) is fulfilled, as
        expected (see table~\ref{tab:title1}). }
   \end{figure}

   Tables~\ref{tab:title1} and \ref{tab:title2} and
   figs.~\ref{fig:confuctivity1} and \ref{fig:confuctivity2} summarize
   the results of applying  our model to well-known experimental data
   \cite{Zukoski1985} for suspensions
   of spherical and nearly monodisperse Latex A ($a= 83 \,{\rm nm}$)
   and Latex B ($a= 235 \,{\rm nm}$) particles in aqueous HCl and KCl
   solutions. The particles were synthesized by two different
   techniques. The concentration $c$ was varied by successive
   dilution of suspension samples. The samples were thermostatted
   to $25 \pm 0.05 ^\circ{\rm C}$. Conductivity measurement results
   with a bridge method at 80 and 1000 Hz were independent of
   frequency. Measurements of the latex electrophoretic  mobility
   were carried out 20-40 times for each salt concentration
   (molarity $M$) and were stable. The $\zeta $-potentials
   (columns 3 in the tables) were determined  using several
   electrokinetic models (\cite{O'Brien1978,OBrien81} and others). The conductivity data
   revealed that for $\kappa a >6 $, the $\sigma_{\rm eff}$ vs $c$
   plots could be considered linear up to $c$ amounting to several
   hundredths.
   The corresponding conductivity increments (columns 4) were
   reported for all data. Several conductivity plots were presented
   for Latex A in HCl.

   \begin{table}[h!]
    \caption {Results of processing $\sigma_{\rm eff}$ vs $c$ data
        \cite{Zukoski1985} for suspensions of latex particles in aqueous
        HCl solutions} \label{tab:title1}
    \begin{tabular}{rrrrrrrrr}
        \hline
        $M$                & $\kappa a$ & $y_\zeta$ & $\cfrac{\partial x_{\rm eff}}{\partial c}$, {\footnotesize exp.} & \multicolumn{1}{l}{$\cfrac{\partial x_{\rm eff}}{\partial c}$, {\footnotesize  calc.}} & $u^*$ & $d^*$, in $\kappa^{-1}$ & (\ref{eq:condIncrease2RadMoment})                      & \multicolumn{1}{l}{T/F} \\    \hline
        \multicolumn{9}{l}{Latex A, $a = 83$ nm}                                                                                  \\
        $1 \times 10^{-2}$ & $26.9$       & $-2.27$     & $-0.31 \pm 0.02$                      & $-0.31 $                                                   & $0.094$ & $2.53$                    & $0.150<0.190$     & T                        \\
        $5 \times 10^{-3}$ & $19.1$       & $-2.33$     & $0.60 \pm 0.05$                       & $0.53$                                                     & $0.153$ & $2.92$                    & $0.249 > -0.033$ & T                        \\
        $1 \times 10^{-3}$ & 8.5        & $-2.44$     & $3.10 \pm 0.01$                       & $3.10$                                                     & $0.278$ & $2.36$                    & $0.006 > -0.196$ & T                        \\
        $5 \times 10^{-4}$ & 6.0        & $-2.17$     & $3.57 \pm 0.06$                       & $3.61$                                                     & $0.280$ & $1.68 $                  & $0.967 > -0.597$ & T                        \\
        \multicolumn{9}{l}{Latex B, $a = 235$ nm}                                                                                                                                                                                               \\
        $5 \times 10^{-3}$ & $54.6$       & $-3.22$     & $-0.85 \pm 0.07$                      & $-0.85 $                                                   & $0.052$ & $2.84$                    & $0.006 < 0.336$     & T                        \\
        $1 \times 10^{-3}$ & $24.4$       & $-3.40$     & $-0.30 \pm 0.15$                      & $-0.30   $                                                 & $0.083$ & $2.03$                    & $0.008 < 0.230 $    & T                        \\
        $5 \times 10^{-4}$ & 17.3       & $-2.89$     & $1.33 \pm 0.08$                       & $1.33$                                                     & $0.202$ & $3.49$                    & $0.007 > -0.237$ & T                        \\
        $1 \times 10^{-4}$ & 7.7        & $-2.28$     & $2.36 \pm 0.23$                       & $2.37$                                                     &$ 0.228$ & $1.76 $                   & $0.004 > -0.352$ & T \\
        \hline
    \end{tabular}
   \end{table}

   Since the values $\kappa a \gtrsim 6$ are expected to belong to
   the region of validity of the Gouy-Chapman solution (\ref{eqGouy}),
   our model is applicable to process the above data. We proceed as
   follows:

   \begin{figure}[bth]
    \centering
    \includegraphics[width=75mm]{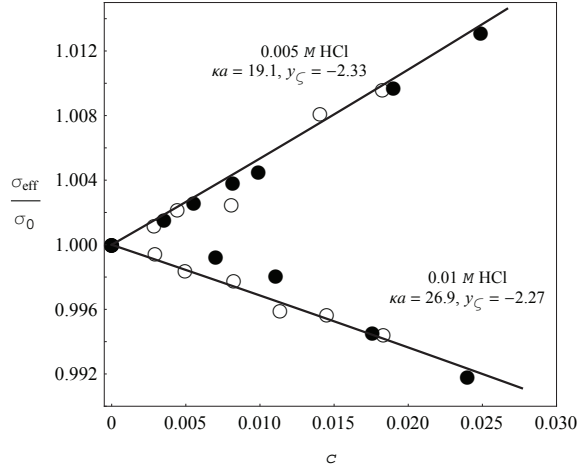}
    \caption{\label{fig:confuctivity1}  $\sigma_{\rm eff}$ as a
        function of $c$ for suspensions of Latex A particles in
        $1\times 10 ^{-2}$ and $5\times 10 ^{-3}$ $M$ HCl; markers $\bullet$ and $\circ$ represent duplicate experiments
        starting from fresh latex
        \cite{Zukoski1985}. Solid lines: our model results for reported $\kappa a$
        and $\zeta$ and the fitting values of $u^* = 0.094$ ($d^*= 2.53\, \kappa^{-1}$) and
        $0.153$ ($d^*=2.92\, \kappa^{-1}$), respectively. The
        corresponding theoretical estimates ($-0.31$ and 0.53) and
        experimental data ($-0.31 \pm 0.02$ and $0.60 \pm 0.05$) for
        $\partial x_{\rm eff}/\partial c $ are in practical agreement.
        Condition (\ref{eq:condIncrease2RadMoment}) is not fulfilled
        and is fulfilled, respectively, as expected (see
        table~\ref{tab:title1}).}
   \end{figure}

   \begin{enumerate}
    \item
    Using given values of $\kappa a$ and $\zeta$-potentials, and taking the ion
    mobilities to be $ \mu_{\rm H^+}= 36.29 \times 10^{-8}\,
    {\rm m ^ 2 / V^{-1} \cdot s^{-1}} $, $\mu_{\rm K^+}=7.6 \times 10^{- 8}\,
    {\rm m ^ 2 / V^{-1} \cdot s^{-1}} $, and $ \mu_{\rm Cl^-}=7.92\times 10^{-8}\,
    {\rm m ^ 2 / V^{-1} \cdot s^{-1}} $, we find that the behavior of
    $x_2(u)$ for latex suspensions in HCl ($m = 4.58$) resembles
    that of curve 1 in fig.~\ref{fig:condtypes}, and that for latex
    suspensions in KCl ($m=0.96$) resembles that of curve 4.
    \item
    Equating the theoretical expressions [(\ref {eq:rate}) and its
    counterparts for the cases of stagnant layers with intermediate and low
    conductivities] and experimental data (columns 4) for the
    suspension conductivity increments, we calculate the values of
    $u^*$ (columns 6) and then verify that they reproduce the reported
    increments within an experimental error (columns 5). The only
    slight exception is the value of $u^*$ for Latex A suspensions in
    $5\times 10 ^{-3}$ $M$ HCl, which is adjusted to fit the entire
    plot of $\sigma_{\rm eff} $ vs $c$ available (see
    fig.~\ref{fig:confuctivity1} ). Nonetheless, the adjusted and
    reported values for $\partial x_{\rm eff}/\partial c$ remain in practical agreement.
    \subitem
    It must be emphasized that physically meaningful
    solutions for $u^*$ occur only in the case of highly conductive
    stagnant layers.
    \item
    Using the values of $u^*$ obtained, we find that the slip
    plane in the suspensions under consideration is located near the
    edge (in the traditional physical meaning) of the diffuse EDL (columns 7). As the ionic strength of the
    electrolyte solutions decreases, this plane seems to have a
    tendency to penetrate deeper into the EDL. This can be explained
    by ``softening'' of the latter.
    \item
    Finally, calculating the values of the left- and right-hand
    sides in (\ref{eq:condIncrease2RadMoment}) and
    (\ref{eq:condDecrease2}), we make sure that in all cases, the
    directions of the resulting inequalities are in line with the
    expected ones (columns 8 and 9). This fact signifies the internal
    consistency of the proposed  model.
   \end{enumerate}

   So, our model, which employs a single fitting parameter $u^*$, is
   capable of describing $\sigma_{\rm eff}$ of the above suspensions
   of nanosized particles in a cohesive manner, the estimated values
   of $u^*$ exhibiting similar trends. These processing results
   indicate that for  the specified values of $\kappa a$, $y_\zeta$,
   and $c$, the electrical conductivity of the considered suspensions
   is contributed to by two factors: the presence of a highly conductive stagnant
   layer inside the slip plane and transport of mobile ions beyond
   it. Simple independent  estimates in support of the existence of
   highly conductive interphase layers in nanofluids can also be found
   elsewhere \cite {Sushko2016}.

   It should also be remarked that rather considerable (in terms of
   the Debye length) values of $u^*$ raise the question of accuracy
   of evaluating the $\zeta $-potential of nanofluids with standard
   electrophoretic models; occurring inaccuracies may necessitate
   adjustments to the above results. A feasible approach to attacking
   this question may consist in a combined analysis of the results
   obtained with the proposed model for $\sigma_{\rm eff}$ and those
   with the method of laser correlation spectroscopy for the
   diffusion coefficient of nanoparticles in the nanofluid under
   study  \cite{Balika2022}.

   \begin{table}[h!]
    \caption {Results of processing $\sigma_{\rm eff}$ vs $c$ data
        \cite{Zukoski1985} for suspensions of latex particles in aqueous
        KCl solutions } \label{tab:title2}
        \begin{tabular}{rrrrrrrrr}
            \hline
            $M$                & $\kappa a$ & $y_\zeta$ & $\cfrac{\partial x_{\rm eff}}{\partial c}$, {\footnotesize  exp.} & \multicolumn{1}{l}{$\cfrac{\partial x_{\rm eff}}{\partial c}$, {\footnotesize  calc.}} & $u^*$ & $d^*$, in $\kappa^{-1}$ & (\ref{eq:condDecrease2})                      & \multicolumn{1}{l}{T/F} \\
            \hline
            \multicolumn{9}{l}{Latex A, $a = 83$ nm}                                                                                                                                                                                                \\
            $1 \times 10^{-2}$ & $26.9 $      & $-3.06$     & $-0.85 \pm 0.05$                      & $-0.85   $                                                 & $0.050$ & $1.35 $                   & $0.018 < 0.114$     & T                        \\
            $1 \times 10^{-3}$ & 8.5        & $-2.96$     & $1.55 \pm 0.08$                       & $1.55 $                                                    &$ 0.209$ & $1.78 $                   & $0.042 > -0.089$ & T                        \\
            \multicolumn{9}{l}{Latex B, $a = 235$ nm}                                                                                                                                                                                               \\
            $5 \times 10^{-3}$ &$ 54.6 $      & $-3.22$     & $-0.57 \pm 0.07$                      & $-0.58   $                                                 & $0.082$ & $4.48 $                   & $0.014 < 0.078$     & T                        \\
            $1 \times 10^{-3}$ & 24.4       & $-3.40$     & $0.00 \pm 0.05$                       & $-0.005$                                                   &$ 0.120$ & $2.93 $                   & $0.028 < 0.032 $    & T \\
            \hline
        \end{tabular}
   \end{table}

  \section{Conclusion} \label{sec5}

  Finding the electrical conductivity $\sigma_{\rm eff}$ of
  colloidal suspensions is a long-standing problem. It is typically
  studied by analyzing a system of coupled nonlinear equations and
  relations describing the ion transport in the vicinity of a single
  particle. Practical advances are possible after using a number of
  poorly controlled  or uncontrolled approximations, employing model
  boundary conditions, and ignoring electromagnetic inter\-actions
  between suspended particles.  The individual roles of the
  parameters of the model are difficult to keep track of.

  Considering a suspension as a system of particles with the hard--core---inhomogeneous--penetrable--shell
  morphology, we propose a new many-particle approach to the problem. As compared to previous publications
  in the field, its distinctive features are as follows:
  (1) it effectively incorporates both many-particle electro\-mag\-ne\-tic interactions among the structural units
  (particles, EDLs, and suspending liquid) of the suspension and variations of the system's micro\-struc\-tu\-re
  as the concentration of the particles is changed;
  (2) the microstructure of the EDL is extended to include the stagnant layer (or an analogue of it) in
  between the Stern layer and the mobile diffuse part of the EDL;
  (3) under the assumptions of the model  (modelling the microstructure of the system in terms of the
  electrical conductivity distributions in its structural units subject to certain rules of dominance for
  overlapping regions, and using our original method of compact groups of inhomogeneities for electrodynamic
  homogenization), the quasistatic $\sigma_{\rm eff}$ is calculated rigorously without in-depth
  approximations typical of one-particle theories;
  (4) the rigorous integral relation obtained for $\sigma_{\rm eff}$ makes it possible to analyze individual
  roles of the parameters of the structural units (such as the ion concentrations and mobilities,  slip
  plane location, $\zeta$-potential, particle volume concentration,
  etc.) on  $\sigma_{\rm eff}$ and to develop a system of inequalities for controlling the internal
  consistency of the model.

  Specific features and efficiency of the model are demonstrated by
  elaborating and applying the general relation to the case of
  latex suspensions in aqueous electrolyte solutions with high ionic
  strength. A single fitting
  parameter, the relative thickness of the stagnant layer, proves to be sufficient
  to describe experiment.

  The effect of the stagnant layer (the region in between the Stern layer and the slip plane) on
  $\sigma_{\rm eff}$ can remain significant in the systems where the mobile  diffuse part of the EDL
  is suppressed.   Based on our model,  we demonstrated this fact recently  \cite{Sushko2022}
  for  concentrated suspensions of so-called ghost particles (fabricated by lysis from human erythrocytes)
  in aqueous NaCl and phosphate-buffered saline
  solutions of molarity 0.15.

  Under certain conditions, both parts of the EDL can  partially contribute  to $\sigma_{\rm eff}$ of
  non-aqueous  systems as well. For instance, the formation of an interface layer is reported based
  on the results of molecular dynamic simulation for Cu particles in liquid Ar
  \cite{Li2010}   and  caloric measurements for suspensions of Al$_2$O$_3$ particles in isopropyl
  alcohol   \cite{Zhelezny2019}.   On the other hand, the formation of the EDL-like layer  around the
  particles can occur in such systems  when the base liquid is slightly contaminated.
  Using     an earlier version of the model under consideration, the effect of this layer on
  $\sigma_{\rm eff}$ was demonstrated in a specially-designed experiment for  suspensions
  of Al$_2$O$_3$ in isopropyl alcohol
  \cite{Sushko2016}.
  It should be noted that in cases where the EDL is of little or no significance, our general
  formalism still applies and can be used to study, for instance, the properties of and the contribution
   to $\sigma_{\rm eff}$ from the suspending liquid \cite{Sushko2019pre}.

  The model   allows for further   diversification   and can be considered as
  a flexible tool for   analysis and design of electrical properties of suspensions.


\end{document}